\documentclass[
prl,
reprint,
superscriptaddress,
showpacs,
amsmath,amssymb,
floatfix,
]{revtex4-1}
\usepackage{dcolumn}
\usepackage{bm}
\usepackage{turnstile}

\usepackage{graphicx}
\usepackage{amsmath,amssymb}
\usepackage{url}
\usepackage{multirow}
\usepackage{float}
\usepackage[toc,page]{appendix}
\usepackage{color}
\DeclareMathOperator{\tr}{\mathop{\mathrm{Tr}}}

\begin{document}

\title{Competitive helical bands and highly efficient diode effect in F/S/TI/S/F hybrid structures}
\author{T.~Karabassov}\email{tkarabasov@hse.ru}
    \affiliation{Moscow Institute of Physics and Technology, Dolgoprudny, Moscow 141700, Russia}
\author{I.~V.~Bobkova}
    \affiliation{Moscow Institute of Physics and Technology, Dolgoprudny, Moscow 141700, Russia}
\author{V.~S.~Stolyarov}
	\affiliation{Moscow Institute of Physics and Technology, Dolgoprudny, Moscow 141700, Russia}
 \affiliation{Dukhov Research Institute of Automatics (VNIIA), 127055 Moscow, Russia}
\author{V.~M.~Silkin}
\affiliation{Donostia International Physics Center (DIPC), San Sebasti\'{a}n/Donostia, 20018 Basque Country, Spain}
\affiliation{Departamento de F\'{\i}sica de Materiales, Facultad de Ciencias Qu\'{\i}micas,
    UPV/EHU, 20080 San Sebasti\'{a}n, Basque Country, Spain}
\affiliation{IKERBASQUE, Basque Foundation for Science, 48011 Bilbao, Spain}
\author{A.~S.~Vasenko}\email{avasenko@hse.ru}
    \affiliation{HSE University, 101000 Moscow, Russia}
    \affiliation{Donostia International Physics Center (DIPC), 20018 San Sebastián-Donostia, Euskadi, Spain}

	\begin{abstract}
        The diode effect in superconducting materials has been actively investigated in recent years. Plenty of different devices have been proposed as a platform to observe the superconducting diode effect. In this work we discuss the possibility of a highly efficient superconducting diode design with controllable polarity. We propose the mesoscopic device that consists of two separated superconducting islands with proximity induced ferromagnetism deposited on top of the three-dimensional topological insulator. Using the quasiclassical formalism of the Usadel equations we demonstrate that the sign of the diode efficiency can be controlled by magnetization tuning of a single superconducting island. Moreover, we show that the diode efficiency can be substantially increased in such device. We argue that the dramatic increase of the diode efficiency is due to competing contribution of the two superconducting islands to the supercurrent with single helical bands linked through the topological insulator surface.
        \end{abstract}
	
	
	\maketitle
	
{\it Introduction.}---
 The superconducting nonreciprocal phenomena has been attracting a lot of attention over the last several years \cite{Nadeem_arxiv}. Particularly, the diode effect in superconducting systems has been widely discussed due to its interesting underlying physics and potential application in the nondissipative superconducting electronics\cite{Soloviev2017,Linder2015,Eschrig2015}. So far the superconducting diode effect has been reported in many different systems, Josephson junctions \cite{Golod2022_arxiv,Wu2022, Baumgartner2022, Pal_arxiv,Chen2018,Trahms_arxiv,yu2024time},including junction-free devices \cite{Ando2020, Narita_arxiv, Itahashi2020,Lin_arxiv,Yasuda2019,teknowijoyo2023flux}, superconducting micro-bridges\cite{suri2022non,chahid2023high} and other systems\cite{Lyu2021,satchell2023supercurrent}. There have been numerous theoretical propositions demonstrating the possibility of superconducting diode effect such as bulk superconducting materials \cite{Scammell_arxiv,Yuan_arxiv,He_arxiv,Daido2022,Ilic_arxiv,Legg2022_arxiv, banerjee2024enhanced,chen2024intrinsic,nakamura2024orbital,hasan2024supercurrent,kubo2023tuning,mironov2024photogalvanic,cadorim2024harnessing}, proximity effect hybrid structures\cite{Devizorova2021,Karabassov2021,Karabassov2022,Karabassov2023,kokkeler2024nonreciprocal,banerjee2024altermagnetic,hosur2023proximity,wang2024secondary,kopasov2022nucleation,seleznyov2024ferromagnetic,neilo2024spin}, Josephson structures \cite{Grein2009,Yokoyama2014,Kopasov2021,Davydova_arxiv,Kokkeler2022,Zazunov2023,lu2023tunable,cayao2024enhancing,seoane2024tuning,wang2024efficient,wei2023josephson,mao2024universal,alvarado2023intrinsic,chatterjee2023quasiparticles,huang2024superconducting,karabassov2024anisotropic,samokhvalov2024anomalous,vakili2024field,fu2024field,guarcello2024efficiency}, nanotubes\cite{he2023supercurrent}, confined systems \cite{de2023superconducting}, asymmetric SQUIDs \cite{fominov2022asymmetric,cuozzo2024microwave,seleznev2024influence} and superconducting systems with nonuniform magnetization \cite{roig2024superconducting}. The diode effect might be useful not only from application point of view, but it may be also employed as a way to detect the spin-orbital coupling (SOC) type of the material\cite{Amundsen2024}.

Typically, such devices require three ingredients for achieving the nonreciprocity of the critical current, including lack of the inversion and time-reversal symmetries and superconductivity\cite{Nadeem_arxiv}. However, it should be emphasized that the lack of inversion symmetry is the implication of the gyrotropy in the structure of the material that supports nonreciprocal transport \cite{kokkeler2024nonreciprocal}. On the microscopic level the lack of inversion symmetry is expressed by the SOC term. In this regard systems based on the topological insulators (TI) are interesting since they offer strongest SOC rendering linear spin-polarized dispersion for the surface states\cite{Hasan2010}.   

The diode effect in the TI based structures has been reported in Josephson junctions, as well as in hybrid structures. In practice, when producing mesoscopic diode devices it is reasonable to expect some presence of the nonmagnetic impurities in the structures. However, previously it has been shown that the diode efficiency is expected to be low in the diffusive TI based systems \cite{Kokkeler2022,Karabassov2022}. Another disadvantage of the TI diffusive diodes is their limited tunability. In these devices the polarity of the diode cannot be changed without reversing the Zeeman field, although in the long ballistic S/TI/S (S denotes superconductor) Josephson junctions such situation is possible\cite{Lu2023}. 

\begin{figure}[t]
	\centering
	\includegraphics[width=\columnwidth]{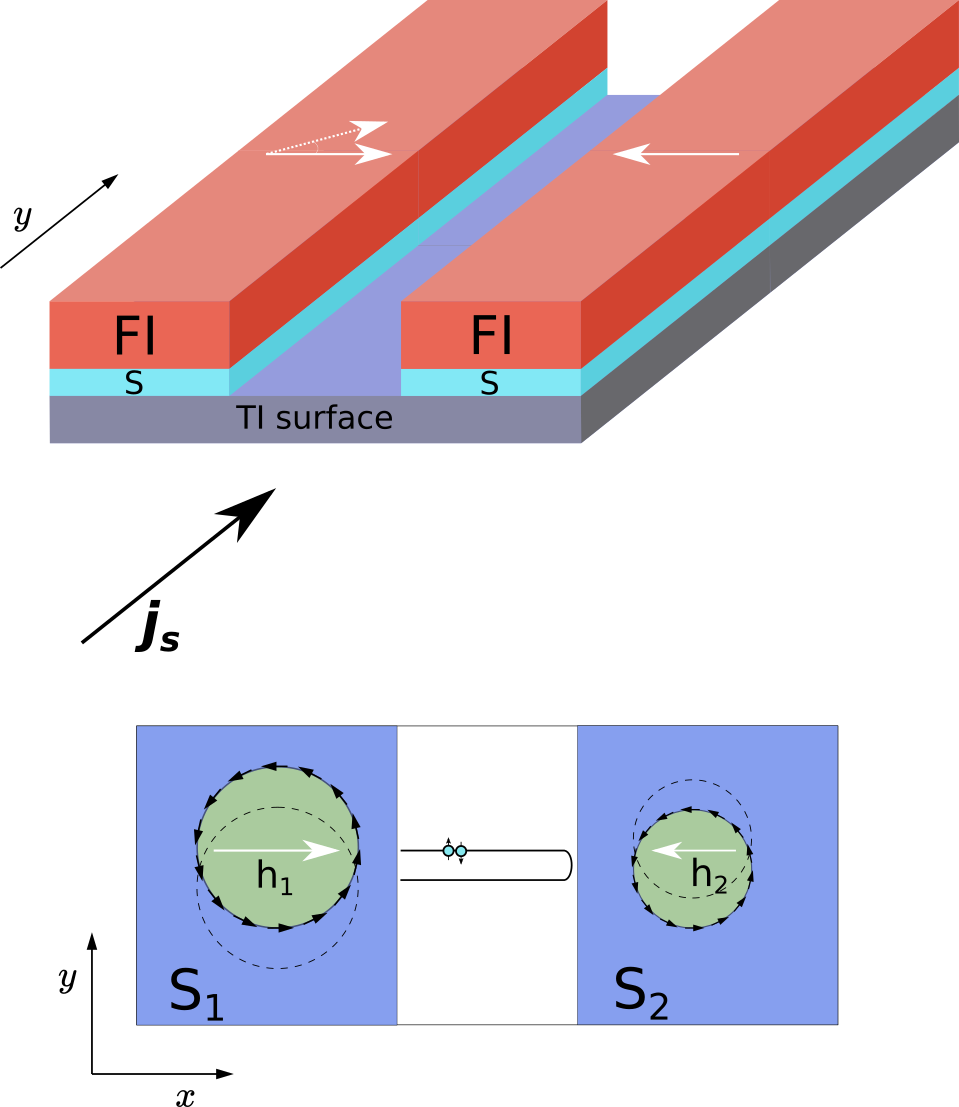}
	\caption{Geometry of the controllable diode under consideration, which consists of two superconducting islands with the proximity induced in-plane exchange field deposited on top of the topological insulator. Schematic representation of the Fermi contours of the two superconducting regions with the exchange fields oriented in the opposite directions. $S_1$ and $S_2$ are linked through the TI surface.}
	\label{model}
\end{figure}

In the present work we propose the superconducting diode based on two superconducting regions with proximity induced in-plane exchange field on top of the topological insulator. The Fermi contour of the TI surface states is usually represented by the Dirac spectrum, i. e. single helical band, which is characterized by the strongest spin-momentum locking effect. Here we consider F/S/TI/S/F (F denotes ferromagnetic layer) hybrid structure, which is depicted in Fig.~\ref{model}. We argue that such hybrid structure can behave as a two helical band system, as for example  noncentrosymmetric superconductors\cite{Houzet2015,Ilic_arxiv}. However, the two helical bands in the structure under consideration are coupled not in the momentum space but in the real space by the TI surface. The coupling between the two islands can be controlled ,for example, by the width of the non-superconducting TI part. When considering the diode effect the proposed layout can substantially increase the diode efficiency, provided the ferromagnetic exchange fields of the two F/S regions are oriented in the opposite directions. Misalignment of the exchange fields leads to the competition of the two separate helical bands in the superconducting regions in their contribution to the critical current nonreciprocity (Fig.\ref{model}).

{\it Quasiclassical theory.}---
 The F/S/TI/S/F hybrid structure can be described by the following effective low-energy Hamiltonian in the particle-hole and spin space
\begin{equation}\label{Ham}
    H (\textbf{k}) = \alpha \left( k_y \sigma_x - k_x \sigma_y\right) \tau_z - \left( \mu + V (\textbf{k})\right)\tau_z + \textbf{h} \cdot \bm \sigma \tau_0 - \hat{\Delta}(\bm k),
\end{equation}
where $\alpha$ is the Fermi velocity, $\mu$ is the chemical potential, $V$ is the impurity potential, $\bm h = (h_x, 0,0)$ is the exchange field due to the adjacent ferromagnetic. The matrices $\tau$ and $\sigma$ are $2\times2$ Pauli matrices in the particle-hole and spin spaces, respectively. The superconducting pair potential matrix $\hat{\Delta}$ is defined as $\hat{\Delta}= i \sigma_y \hat{\Delta}_s =i \sigma_y \hat{U} i \tau_x \Delta (x) \hat{U}^\dagger$, where transformation matrix $\hat{U}= \exp \left( i q y \tau_z/2 \right)$. The finite center of mass momentum $q$ takes into account the helical state. Pair potential $\Delta(x)$ is a real function defined in the following way
\begin{equation}
    \Delta(x) = \begin{cases}
        \Delta_1 (x),\quad -d_{s1} - L/2<x< -L/2  \\
        0,\quad - L/2<x< L/2 \\
        \Delta_2 (x),\quad L/2<x< d_{s2} + L/2.
    \end{cases}
\end{equation}
Here $\Delta_1$ and $\Delta_2$  are calculated self-consistently and correspond to the superconducting regions $S_1$ and $S_2$, respectively (Fig. \ref{model}). Finally, $L$ is the width of the bare TI surface (normal N part) and $d_{s1}$($d_{s2}$) is the width of $S_1$ ($S_2$) region. It is significant to emphasize that although the geometry of the considered device corresponds to the Josephson junction, in this work we consider zero macroscopic phase difference between regions $S_1$ and $S_2$, so that the Josephson supercurrent is absent. The anomalous ground state phase shift $\phi_0$ is also absent since we assume exchange field component $h_y=0$. On the other hand $h_x$ component is considered to be finite in the system and defined as follows
\begin{equation}
    h_x = \begin{cases}
        h_1,\quad -d_{s1} - L/2<x< -L/2  \\
        0,\quad - L/2<x< L/2 \\
       h_2,\quad L/2<x< d_{s2} + L/2.
    \end{cases}
\end{equation}

We solve the stated problem for Hamiltonian in Eq. \eqref{Ham} within the microscopic approach based on the quasiclassical Green's functions in the diffusive limit. Such model can be described by the Usadel equations \cite{Zyuzin2016,Bobkova2017,ozaeta2012}

\begin{equation}\label{Usadel_general}
D \hat{\nabla}\left(\hat{g} \hat{\nabla} \hat{g} \right)= \left[\omega_n \tau_z + i \hat{\Delta}_s, \hat{g}\right].
\end{equation}
Here $D$ is the diffusion constant, $\tau_z$ is the Pauli matrix in the particle-hole space. In general case the operator $\hat{\nabla} X = \nabla X + i \left(h_x \hat{e}_y - h_y \hat{e}_x\right) \left[\tau_z, \hat{g}\right]/\alpha$.  The Green's function matrix is also transformed as $\hat{g}= \hat{U} \hat{g}_q \hat{U}^\dagger$.

To facilitate the solution procedures of the nonlinear Usadel equations we employ $\theta$ parametrization of the Green's functions\cite{Belzig1999},
\begin{equation}
\hat{g}_q= 
\begin{pmatrix}
\cos \theta & \sin \theta \\
\sin \theta & -\cos \theta
\end{pmatrix}.
\end{equation}
Substituting the above matrix into the Usadel equation \eqref{Usadel_general}, we obtain in the superconducting S parts $|x|> L/2$:
\begin{align}
\xi_{s}^2 \pi T_{cs} & \left[ \partial_x^2 \theta_{i}^s - \frac{q_i^2 }{2} \sin 2 \theta_{i}^s \right]= \\
& =\omega_n \sin{\theta_{i}^s} - \Delta_i (x) \cos{\theta_{i}^s}, \nonumber
\end{align}
where index $i=1,2$ refers to the superconducting parts $S_1$ and $S_2$, $q_i = q + 2 h_i /\alpha$ and in the normal N part $-L/2<x<L/2$:
\begin{align}
\xi_{N}^2 \pi T_{cs} & \left[ \partial_x^2 \theta_{N} - \frac{q^2 }{2} \sin 2 \theta_N \right] =\omega_n \sin{\theta_{N}},
\end{align}
where $\theta_{s(N)}$ means the value of $\theta$ is the S(N) of the TI surface, respectively. We introduced the characteristic length $\xi_{s(N)} = \sqrt{D_{s(N)}/ 2 \pi T_{cs}}$, where $T_{cs}$ is the transition temperature of the bare S region. The self-consistency equations for the pair potentials read,
\begin{equation}
\Delta_i (x) \ln \frac{T_{cs}}{T} = 2 \pi T \sum_{\omega_n>0} \left( \frac{\Delta_i (x)}{\omega_n} - \sin \theta_i^s \right).
\end{equation}
Finally we supplement the above equations with two pairs of the boundary conditions (two for each S/N interface) of the following type
\begin{eqnarray}
    \gamma \xi_l \hat{g}_l {\nabla}\hat{g}_l = \xi_r \hat{g}_r {\nabla}\hat{g}_r, \\
    \gamma_{B} \xi_l \hat{g}_l {\nabla}\hat{g}_l = [\hat{g}_l,\hat{g}_r].
\end{eqnarray}
Here parameters $\gamma_B = R_B \sigma_l / \xi_l$, $\gamma = \xi_r \sigma_l/ \xi_l \sigma_r$ where $\sigma_{l(r)}$ is the conductivity of the material on the left (right) of the interface. Parameter $\gamma$ controls the slope of the Green's functions at the interface, whereas $\gamma_B$ controls the mismatch between the functions at the interface. While for identical materials $\gamma = 1$, in general this parameter may have arbitrary value. $\gamma_B$ is the parameter that determines the transparency of the S/F interface \cite{KL, VB1, VB2}.

\begin{figure}[t]
	\centering
	\includegraphics[width=\columnwidth]{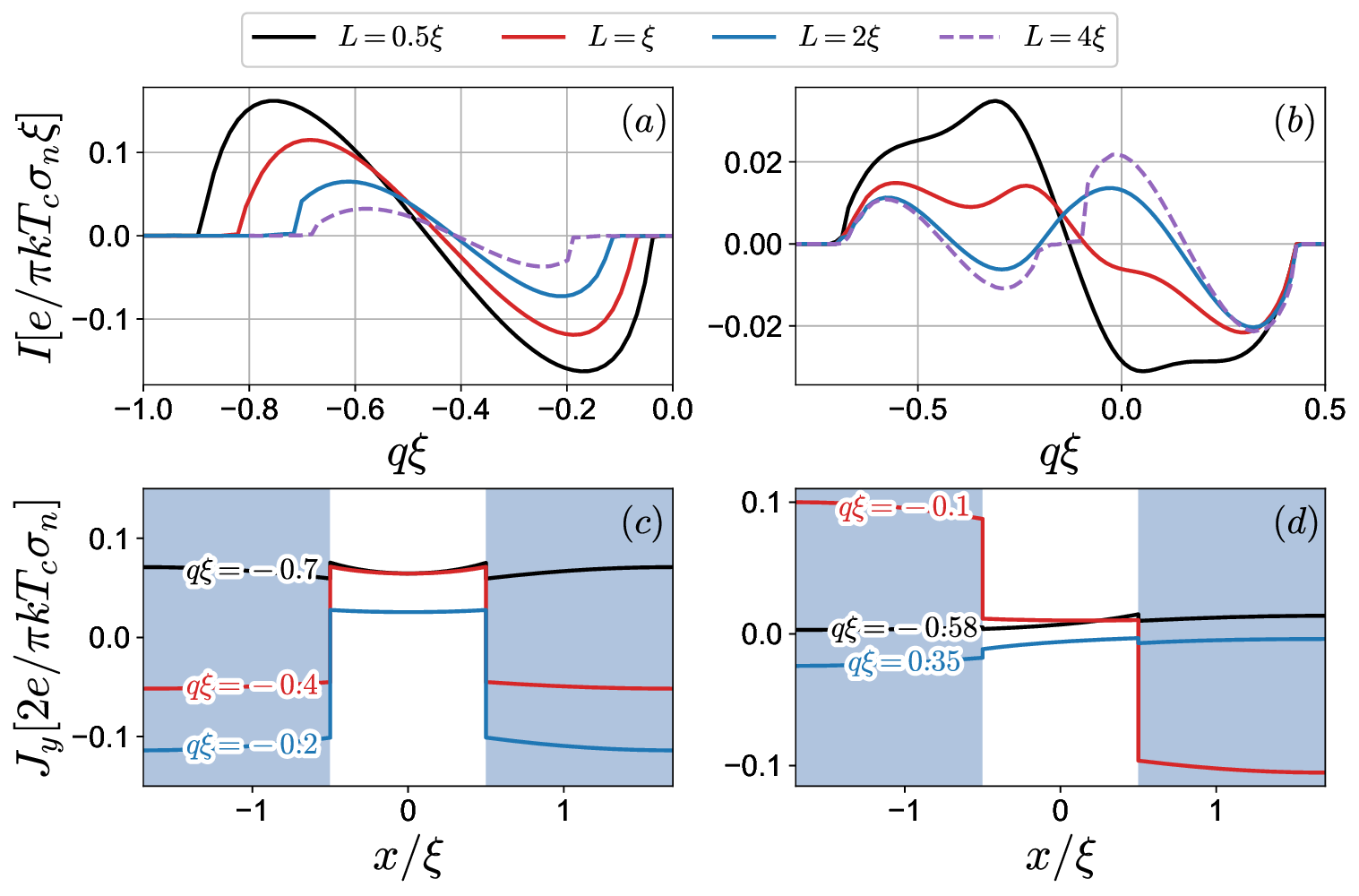}
	\caption{Supercurrent $I$ as a function of $q$ at $h_1 = h_2 = 0.25$ (a) and at $h_1 = -0.1, h_2 = 0.25$ (b). Lower panels illustrate the current density distributions at different $q$ corresponding to the upper panels for $L = \xi $.}
	\label{fig1}
\end{figure}

The supercurrent in the diffusive limit can be found from the following expression 
\begin{equation}
\textbf{J}_{s(N)}= \frac{- i \pi \sigma_{s(N)}}{4 e} T \sum_{\omega_n} \tr \left[ \tau_z \hat{g}_{s(N)} \hat{\nabla} \hat{g}_{s(N)} \right].
\end{equation}
Performing the unitary transformation $U$, the current density transforms as follows:
\begin{align}
{j}_y^{s1(2)} (x)=- \frac{\pi \sigma_{s1(2)}  }{2 e} \left[ q  + \frac{2 h_{1(2)}}{\alpha}\right] T \sum_{\omega_n} \sin^2 \theta_{s1(2)}, \\
{j}_y^n (x)=- \frac{\pi \sigma_n  q }{2 e} T  \sum_{\omega_n} \sin^2 \theta_n.
\end{align}
The total supercurrent flowing via the system along the $y$-direction can be calculated by integrated the current density of the total width of the F/S/TI/S/F:
\begin{align}\label{I_total}
I = I_{s1} + I_{s2} + I_{N},
\end{align}
where $I_{s1}$ , $I_{s2}$ and $I_{N}$ are the total supercurrents integrated along the $x$ direction in $S_1$, $S_2$ and N regions, respectively. 

{\it Results.}---

We fix the following system parameters throughout the discussion of the results: $d_{s1}= d_{s2} = 1.2 \xi$, $\gamma_1 = \gamma_2 = 0.5$, $T=0.1 T_{cs}$. 

We start with the analysis of the $I(q)$ relations when the exchange fields $H_1$ and $H_2$ are the same in both superconducting regions. In Fig. \ref{fig1} (a) we observe a characteristic behavior of the supercurrent with $I(q_0)=0$, where $q_0\ne 0$ is the ground state Cooper pair momentum, which reflects the helical nature of the superconducting ground state. We can also notice some nonreciprocity of the supercurrent, i. e. $I_c^+ \ne I_c^-$ that is a consequence of the helical state. As we will see below the diode efficiency is quite low and in this case does not exceed several percents. In the absence of any exchange field $I(q=0)=0$, which means that the ground state is conventional state with zero Cooper pair momentum. To get more insight we plot the supercurrent density $J_y$ in Fig.\ref{fig1} (b). Hence, in the situation when $H_1$ and $H_2$ are perfectly aligned we expect well-known behavior of the total supercurrent. 

\begin{figure}[t]
	\centering
	\includegraphics[width=\columnwidth]{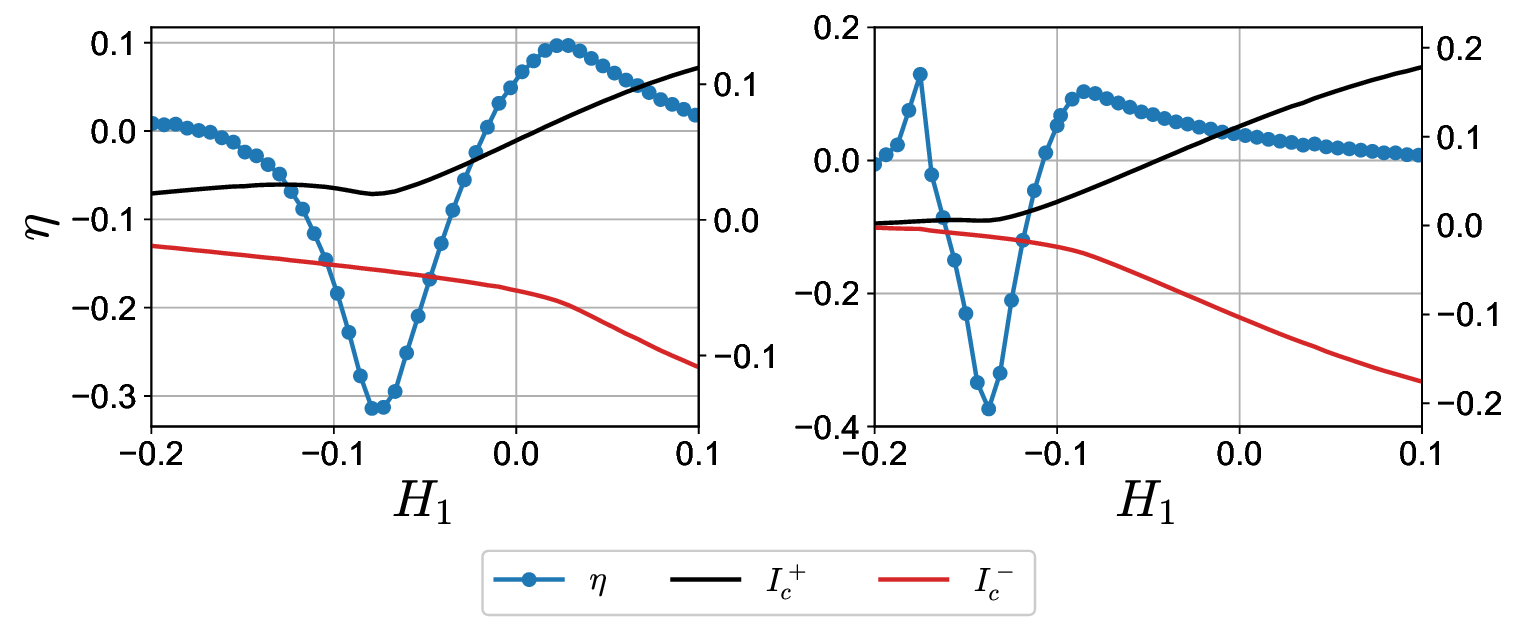}
	\caption{Critical supercurrents $I_c^+$, $I_c^-$ (right vertical scale) and diode efficiency $\eta$ (left vertical scale) as functions of $H_1$ for $L=\xi$. Left and right plots correspond to $\gamma_{B1} = \gamma_{B2} = 0.4$ and $\gamma_{B1} = \gamma_{B2} = 0.2$ respectively. The critical currents scale is in units of $2 e/ \pi k T_c \sigma_n \xi$.}
	\label{fig2}
\end{figure}
Now we discuss the case when the exchange fields $H_1$ and $H_2$  are oriented in the opposite directions (Fig. \ref{fig1} (b)).  When the distance between $S_1$ and $S_2$ is large ($L= 4 \xi$), the superconducting regions are well separated and act as almost independently with distinct critical supercurrents $I_{c1}^{\pm}$ and $I_{c2}^{\pm}$ corresponding to $S_1$ and $S_2$ respectively. This circumstance can be clearly seen from $I(q)$ dependence for $L= 4 \xi$ and in this sense distance $L$ can be imagined as a coupling strength between $S_1$ and $S_2$ . The behavior of $I(q)$ dramatically changes if $L$ becomes smaller. The regions of $I(q)$ curve which previously could be easily assigned to each superconducting island start to "overlap" reflecting stronger coupling between $S_1$ and $S_2$. As a result we can achieve a situation when the critical current of the hybrid structure in one direction is substantially renormalized. For instance, we can observe that $I_c^+$  is defined by rather left maximum of $I(q)$ at $L= \xi$, while $I_c^-$  remains approximately at the same value. Stronger coupling between the superconducting regions leads to a more complicated supercurrent density distribution across the hybrid structure (see Fig. \ref{fig1} (d)). Obtaining such nontrivial behavior of  $I(q)$ is the key idea behind achieving larger diode efficiency $\eta$. It should be emphasized that similar behavior is expected in the Rashba superconductors, where the Fermi surface is represented by the two helical bands with the opposite helicities\cite{Ilic_arxiv,He_arxiv,Daido2022}. Here, we clearly consider a single helical band Fermi surface. However, we can have $S_1$ and $S_2$ with the opposite $h_1$ and $h_2$ in our system  as illustrated in Fig.\ref{model} which may be thought as an effective two helical bands system.

\begin{figure}[t]
	\centering
	\includegraphics[width=\columnwidth]{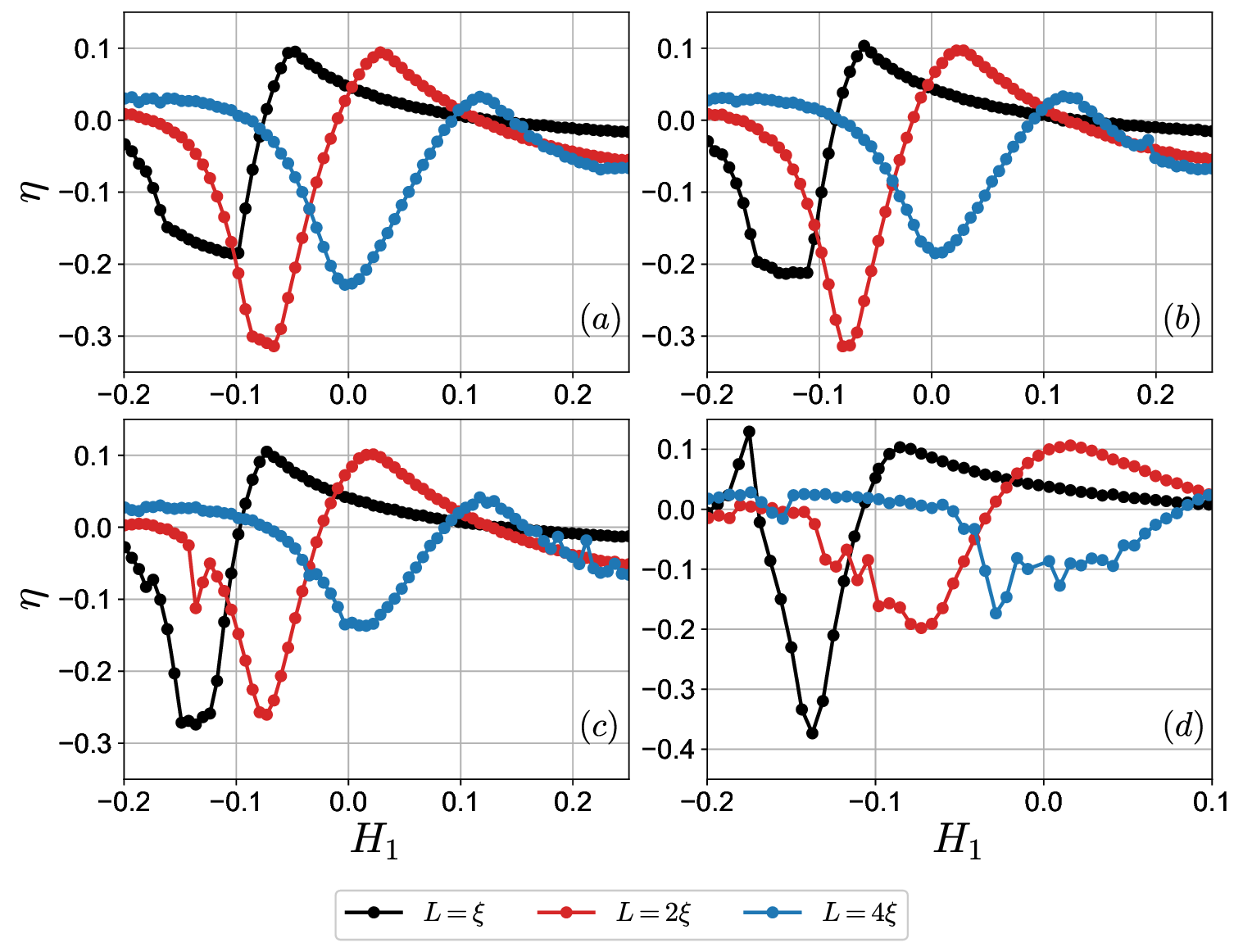}
	\caption{ Superconducting diode efficiency $\eta$ calculated at different interface transparencies $\gamma_B$. Plot (a) corresponds to $\gamma_{B1} = \gamma_{B2} = 0.5$, (b) - $\gamma_{B1} = \gamma_{B2} = 0.4$, (c) - $\gamma_{B1} = \gamma_{B2} = 0.3$ and (d) - $\gamma_{B1} = \gamma_{B2} = 0.2$.}
	\label{fig3}
\end{figure}

The diode efficiency can be defined in a standard way, as
\begin{equation}
    \eta = \frac{I_c^+ - |I_c^-|}{I_c^+ + |I_c^-|}.
\end{equation}
In Fig. \ref{fig2}  the diode efficiency along with the critical currents is demonstrated as a function of $H_1$ , while $H_2$ is fixed at $H_2 = 0.25$. From the figure we observe several characteristic features of $\eta$ behavior. Firstly, the diode efficiency is quite low at large positive values of $H_1$, remaining under 5$\%$ at $H_1 = 0.1$. This is anticipated behavior of the diodes with single helical band in the diffusive limit \cite{Karabassov2022,Karabassov2024_phase,Kokkeler2022}. As $H_1$ decreases the diode efficiency rises to certain value and then $\eta$ changes sign rapidly reaching the maximum value. At the point when the diode changes its polarity, there is a transition between $S_1$ and $S_2$ in their contribution to the critical currents. We assume that in the vicinity of $\eta = 0$ the superconducting regions $S_1$ and $S_2$ strongly compete with each other, since individually they have opposite efficiencies because of $H_1$ and $H_2$ are of the opposite signs.  We might say that at certain value of $H_1$ the critical currents $I_c^+$ and $I_c^-$  of the total system are predominantly determined by $S_1$ and $S_2$, i. e. the supercurrent mostly passes through one of the superconducting regions in the opposite directions. Another important observation from Fig. \ref{fig2} is that the sign change of the diode efficiency occurs at lower values of the critical currents. That means that higher diode efficiencies due to the competition of $S_1$ and $S_2$ take place at substantially suppressed superconducting state. Finally, we can see how the interface transparency affects $\eta$. Higher transparency can increase the efficiency up to $40 \%$ , however at smaller critical currents.

The interface transparency $\gamma_B$ is an important parameter of the system which, in principle, can be used as a tuning parameter in the experiment. Control of this parameter may be achieved by applying the gating voltage at the interface. We provide more detailed analysis of the interface transparency impact on the diode effect in Fig. \ref{fig3}. We notice that the highest efficiency is achieved at smaller $\gamma_B = 0.2$ for $L=\xi$. However, this is not the general trend as we see from the plots. For instance, the highest $\eta$ is realized at $\gamma_B =0.5$ for $L=4 \xi$. Hence, there exists an optimal value of the interface transparency for the highest efficiency. It is also important to emphasize that the exchange field $H_1$ at which 'major' sign change of $\eta$ occurs, shifts towards larger values as $\gamma_B$ decreases. This means that the polarity of the diode can be altered via the control of the interface transparency, which cannot be achieved in the diffusive single helical band superconducting diode \cite{Karabassov2022}. Finally, we observe multiple sign-changing behavior of the quality factor in Fig. \ref{fig3}.  This may reflect the competitive nature of the $S_1$ and $S_2$ behavior in the nonreciprocal supercurrent.

In conclusion, we have examined the superconducting diode effect in the F/S/TI/S/F hybrid structure. It has been shown that at certain condition when the exchange fields of the ferromagnetic regions are opposite the diode efficiency can be dramatically increased. Such improvement can be explained in terms of the competitive behavior of the superconducting regions with single helical bands. The obtained results can be useful for achieving high efficiency superconducting diodes in the absence of the external magnetic field. Moreover, the sign of the diode efficiency can be changed as a function of the interface transparency.

As a direction for further studies one could investigate the Josephson diode effect in the hybrid structure considered in this paper. In this case the nonreciprocity is achieved in the Josephson critical current. 

{\it Acknowledgements.}---
This work is supported by the megagrant of the Ministry of Science and Higher Education of Russian Federation No. 075-15-2024-632. The numerical calculation of the diode efficiency was supported by the Foundation for the Advancement of Theoretical Physics and Mathematics “BASIS” grant number 22-1-5-105-1.

\bibliography{diode.bib}
\end{document}